\documentclass[a4paper]{article}
\usepackage[T1]{fontenc}
\usepackage[utf8]{inputenc}
\usepackage{varwidth}
\usepackage{amsmath}
\usepackage{amsthm}

\makeatletter

\newcommand*\LyXZeroWidthSpace{\hspace{0pt}}
\providecommand{\tabularnewline}{\\}
\newenvironment{cellvarwidth}[1][t]
    {\begin{varwidth}[#1]{\linewidth}}
    {\@finalstrut\@arstrutbox\end{varwidth}}

\numberwithin{equation}{section}
\numberwithin{figure}{section}

\@ifundefined{date}{}{\date{}}

\usepackage{color}
\usepackage{amsfonts}\usepackage{amsthm}\usepackage{amscd}\numberwithin{equation}{section}
\usepackage{bm}
\usepackage[all]{xy}

\usepackage{algorithm}
\usepackage{algorithmicx}
\usepackage{algpseudocode}
\usepackage[most]{tcolorbox}
\tcbuselibrary{listingsutf8}
\usepackage{float}
\usepackage{lipsum}  
\tcbset{
  highlight algorithm/.style={
    colback=gray!10,
    colframe=black,
    boxrule=0.5pt,
    arc=2pt,
    left=6pt,
    right=6pt,
    top=6pt,
    bottom=6pt,
    width=\textwidth,
    sharp corners,
  }
}

\makeatother

\usepackage[english]{babel}
\begin{document}
\textbf{OUJ-FTC-}21\\
\vspace{16pt}

\begin{center}
{\LARGE Supersymmetric M2-Brane Matrix Model with Restricted Volume-Preserving
Deformations:}\\
{\LARGE{} Lorentz Covariance and BPS Spectrum}{\LARGE\par}
\par\end{center}

\begin{center}
\vspace{16pt}
\par\end{center}

\begin{center}
So Katagiri\footnote{So.Katagiri@gmail.com}
\par\end{center}

\begin{center}
\textit{Nature and Environment, Faculty of Liberal Arts, The Open
University of Japan, Chiba 261-8586, Japan} 
\par\end{center}

\begin{center}
\vspace{16pt}
\par\end{center}
\begin{abstract}
We present a novel supersymmetric Lorentz-covariant matrix model for
M2-branes in M-theory, constructed via the framework of Restricted
Volume-Preserving Deformations (RVPD) which is a residual symmetry
from gauge-fixed volume-preserving diffeomorphisms. The model reformulates
the supermembrane action entirely in terms of Nambu brackets, whose
decomposition into Poisson brackets allows for a consistent matrix
regularization that preserves both the Fundamental Identity and Lorentz
covariance. Under the RVPD gauge structure, $\kappa$-symmetry is
reduced to a restricted form ($\tilde{\kappa}$) that closes with
RVPD transformations into a novel symmetry algebra. This algebra enables
a systematic classification of BPS configurations: particle states
(1/2-BPS), noncommutative membranes (1/4-BPS), and extended configurations
in 4, 6, and 8 dimensions (1/8-, 1/16-, and 1/32-BPS), while the 10-dimensional
case is non-BPS. The construction provides a natural lift of D2-brane
matrix theory to M-theory and offers a pathway toward Lorentz-covariant
matrix models for higher branes such as M5-branes.
\end{abstract}

\section{Introduction}

The nonperturbative formulation of M-theory has long been considered
the ``holy grail'' of string theory research\cite{Witten_1995}. In
this quest, matrix models have played a central role.

\LyXZeroWidthSpace{}

The BFSS model\cite{Banks_1997} was proposed as a strong candidate
for a nonperturbative definition of M-theory in the infinite momentum
frame and marked a major breakthrough in this direction. However,
the formulation relies fundamentally on the light-cone gauge, which
explicitly breaks Lorentz invariance. As a result, a manifestly Lorentz
covariant formulation remains an open problem\cite{Fujikawa_1997,Awata_1998,https://doi.org/10.48550/arxiv.hep-th/9906248,Smolin_2000,https://doi.org/10.48550/arxiv.hep-th/0009131,Yoneya_2016,Ashwinkumar_2021}.
Previous approaches to Lorentz covariant matrix models for membranes
exhibit a number of limitations. Fujikawa’s formulation employed BRST
gauge fixing\cite{Fujikawa_1997}, but the resulting expression depends
explicitly on the basis introduced for matrix regularization. Smolin’s\cite{Smolin_2000}
and Minic’s\cite{https://doi.org/10.48550/arxiv.hep-th/0009131} proposals
pointed out the possibility of matrix regularization of the Nambu
bracket, but did not provide a concrete constructive realization.
The matrix model of Awata and Yoneya et al.\cite{https://doi.org/10.48550/arxiv.hep-th/9906248,Yoneya_2016}
satisfy the Fundamental Identity (F.I.) but does not admit a well-defined
classical continuum limit. Ashwinkumar’s regularization of the membrane
algebra\cite{Ashwinkumar_2021} makes use of the Heisenberg--Nambu
bracket, but fails to satisfy the F.I. in general. These limitations
motivate the search for an alternative formulation that preserves
Lorentz covariance, respects the F.I., and admits a consistent classical
limit\footnote{See also \cite{Katagiri_2023} for an attempt to quantize the Nambu
bracket from the operator formalism of classical mechanics.}. 

\LyXZeroWidthSpace{}

In our recent work\cite{https://doi.org/10.48550/arxiv.2504.05940},
we proposed a Lorentz covariant matrix model for bosonic M2-branes
based on \textsl{Restricted Volume-Preserving Deformations} (RVPD).
RVPD arises as a residual symmetry obtained by imposing specific gauge
constraints on the original volume-preserving deformations (VPD).
Concretely, by restricting the behavior of fields along the third
worldvolume coordinate direction ($\sigma^{3}$), one selects a closed
subalgebra within the full VPD symmetry. This construction enables
the avoidance of violations of the Leibniz rule and the Fundamental
Identity---both of which are known obstacles in the matrix regularization
of the Nambu bracket---thus providing a theoretically consistent
framework.

For longstanding efforts addressing the difficulties in matrix regularization
of the Nambu bracket, see \cite{Nambu_1973,Takhtajan_1994,Dito_1997,Sakakibara_2000}.

\LyXZeroWidthSpace{}

In this study, we extend the gauge constraint structure to supersymmetric
membranes and investigate how $\kappa$-symmetry is modified in this
context. In particular, we show that the restricted $\kappa$-symmetry
($\tilde{\kappa}$) forms a closed algebra with the RVPD transformations,
and that this algebraic structure enables a systematic classification
of BPS states. Although RVPD emerges as a byproduct of gauge fixing,
we demonstrate that its algebraic structure carries physical significance---specifically,
it remains consistent with $\kappa$-symmetry and forms a closed symmetry
algebra.

\LyXZeroWidthSpace{}

In summary, our Lorentz-covariant M2-brane matrix model differs fundamentally
from the BFSS model in its gauge structure, symmetry algebra, and
the scope of BPS classification.

A concise comparison is provided in Table \ref{tab:Key-differences-between},
which clearly shows that our approach incorporates RVPD gauge symmetry
with a Nambu 3-bracket structure and achieves a systematic classification
of BPS states beyond what is available in BFSS.

\LyXZeroWidthSpace{}

The structure of this paper is as follows. In Section 2, we provide
an overview of the supermembrane action. In Section 3, we reformulate
the action using Nambu brackets. Section 4 presents a decomposition
of the Nambu bracket into Poisson brackets. In Section 5, we examine
how the RVPD gauge constraint modifies $\kappa$-symmetry and analyze
the structure of its restricted form. In Section 6, we investigate
the BPS stability of noncommutative membrane solutions using the restricted
$\kappa$-symmetry $\tilde{\kappa}$. Finally, Section 7 offers concluding
remarks and future perspectives.

\begin{table}
\centering
\begin{tabular}{|c|c|c|}
\hline 
Aspect & BFSS (1997) & This Work (RVPD + restricted $\kappa$-symmetry)\tabularnewline
\hline 
\hline 
Gauge choice & \begin{cellvarwidth}[t]
\centering
Light-cone gauge 

(Lorentz non-covariant)
\end{cellvarwidth} & \begin{cellvarwidth}[t]
\centering
Lorentz covariant 

(RVPD-preserving)
\end{cellvarwidth}\tabularnewline
\hline 
\begin{cellvarwidth}[t]
\centering
Underlying algebra 

/ gauge structure
\end{cellvarwidth} & \begin{cellvarwidth}[t]
\centering
Standard 2-bracket

(matrix commutator) 

with U(N) gauge symmetry
\end{cellvarwidth} & \begin{cellvarwidth}[t]
\centering
RVPD gauge symmetry 

with Nambu 3-bracket structure 

(decomposed into Poisson brackets

preserving the Fundamental Identity)
\end{cellvarwidth}\tabularnewline
\hline 
$\kappa$-symmetry & \begin{cellvarwidth}[t]
\centering
Not manifest \LyXZeroWidthSpace{}

(gauge-fixed away)
\end{cellvarwidth} & \begin{cellvarwidth}[t]
\centering
Restricted $\kappa$-symmetry $\tilde{\kappa}$

forming closed algebra with RVPD
\end{cellvarwidth}\tabularnewline
\hline 
BPS classification & \begin{cellvarwidth}[t]
\centering
Mostly particle-like 1/2-BPS;

case-by-case for others
\end{cellvarwidth} & \begin{cellvarwidth}[t]
\centering
Systematic: 1/2, 1/4, 1/8, 1/16, 1/32, 

non-BPS
\end{cellvarwidth}\tabularnewline
\hline 
\begin{cellvarwidth}[t]
\centering
Higher-dimensional 

noncommutative membranes
\end{cellvarwidth} & \begin{cellvarwidth}[t]
\centering
Appears via dimensional reduction 

or background fields
\end{cellvarwidth} & \begin{cellvarwidth}[t]
\centering
Direct mapping:$[X{{}^1},X{{}^2}]=i$

\ensuremath{\leftrightarrow} noncommutative D2 gauge field
\end{cellvarwidth}\tabularnewline
\hline 
Novelty & \begin{cellvarwidth}[t]
\centering
Nonperturbative M-theory proposal

in light-cone frame
\end{cellvarwidth} & \begin{cellvarwidth}[t]
\centering
Lorentz-covariant matrix model

with unified BPS spectrum

and high-dimensional extensions
\end{cellvarwidth}\tabularnewline
\hline 
\end{tabular}

\caption{Key differences between BFSS and our Lorentz-covariant matrix model\label{tab:Key-differences-between}}

\end{table}

\section{Review of RVPD}

In this section, we briefly review the Restricted Volume-Preserving
Deformation (RVPD) that we proposed in our previous work\cite{https://doi.org/10.48550/arxiv.2504.05940}.

We consider the membrane action of the form\footnote{For the bosonic case, various approaches to reformulating the Nambu--Goto
action in terms of an n-bracket and gauge theory form have been discussed
in earlier literature\cite{Sugamoto_1983}. A particularly explicit
implementation, removing the square root and recasting the action
as a Filippov--Lie algebra gauge theory, was given by Park and Sochichiu\cite{Park_2009}. }
\begin{equation}
S=\int d^{3}\sigma\frac{1}{2}\{X^{I},X^{J},X^{K}\}^{2},
\end{equation}

where $X^{I}(\sigma^{1},\sigma^{2},\sigma^{3})$ represents the spacetime
coordinates of the membrane with $I=0,1,\dots,10$, and $\sigma^{i}$
(for $i=1,2,3$) are the internal parameters on the worldvolume of
the membrane.

The expression $\{X^{I},X^{J},X^{K}\}$ denotes the Nambu bracket,
which is defined as

\begin{equation}
\{X^{I},X^{J},X^{K}\}=\epsilon^{ijk}\frac{\partial X^{I}}{\partial\sigma^{i}}\frac{\partial X^{J}}{\partial\sigma^{j}}\frac{\partial X^{K}}{\partial\sigma^{k}}.
\end{equation}

The membrane action is invariant under volume-preserving deformations
(VPD), which can be expressed using arbitrary charges $Q_{1}$ and
$Q_{2}$ as
\begin{equation}
\delta X^{I}=\{Q_{1},Q_{2},X^{I}\}.
\end{equation}

This invariance suggests that, if the Nambu bracket could be quantized
analogously to the Poisson bracket, one might obtain a Lorentz covariant
matrix regularization of the membrane. However, no known method allows
for such a quantization that preserves the Fundamental Identity (F.I.)---which
encodes the consistency of ternary operations---within a three-algebra
framework.

\LyXZeroWidthSpace{}

In our previous work\cite{https://doi.org/10.48550/arxiv.2504.05940},
we decomposed the Nambu bracket as
\begin{equation}
\{X^{I},X^{J},X^{K}\}=\{\tau(X^{I},X^{J}),X^{K}\}+\frac{\partial X^{K}}{\partial\sigma^{3}}\{X^{I},X^{J}\}+\Sigma(X^{I},X^{J};X^{K}),
\end{equation}

where
\begin{equation}
\tau(A,B)\equiv\frac{\partial A}{\partial\sigma^{3}}B-\frac{\partial B}{\partial\sigma^{3}}A,
\end{equation}

\begin{equation}
\Sigma(X^{I},X^{J};X^{K})\equiv A\{\frac{\partial B}{\partial\sigma^{3}},C\}-B\{\frac{\partial A}{\partial\sigma^{3}},C\},
\end{equation}

and

\begin{equation}
\{A,B\}\equiv\epsilon^{ab}\frac{\partial A}{\partial\sigma^{a}}\frac{\partial B}{\partial\sigma^{b}},\ a,b=1,2.
\end{equation}

We then imposed the following restrictions on the VPD charges $Q_{1}$
and $Q_{2}$, thereby defining the \textsl{Restricted Volume-Preserving
Deformations} (RVPD):

\begin{equation}
\frac{\partial\tau(Q_{1},Q_{2})}{\partial\sigma^{3}}=0,\ \{Q_{1},Q_{2}\}=0,\ \frac{\partial}{\partial\sigma^{3}}\frac{\partial Q_{1,2}}{\partial\sigma^{a}}=0,\ a=1,2.
\end{equation}

These conditions arise from imposing the gauge constraint
\begin{equation}
C_{I}\frac{\partial X^{I}}{\partial\sigma^{3}}=\sigma^{3},
\end{equation}

as demonstrated in Appendix B of \cite{https://doi.org/10.48550/arxiv.2504.05940}.

\LyXZeroWidthSpace{}

With this decomposition, the resulting matrix-regularized model maintains
the properties associated with the Fundamental Identity (F.I.) even
under continuous RVPD transformations. As such, it provides a consistent
bosonic matrix model for M2-branes without relying on a light-cone
frame. The model admits a variety of solutions, including particle-like
configurations and noncommutative membranes in 2, 4, 6, 8, and 10
dimensions.

\LyXZeroWidthSpace{}

In the next section, we investigate whether these solutions remain
stable under supersymmetry by analyzing their BPS conditions.

\section{Super M2-Brane}

The action for the supersymmetric M2-brane was given in \cite{Bergshoeff_1987}
as
\begin{equation}
S=S_{NB}+S_{WZ},
\end{equation}

where
\begin{equation}
S_{NB}=-\frac{T}{2}\int d^{3}\sigma\left(\frac{1}{e}\left(\epsilon^{ijk}\Pi_{i}^{I}\Pi_{j}^{J}\Pi_{k}^{K}\right)^{2}+e\right),
\end{equation}

\begin{equation}
S_{WZ}=\frac{i}{3!}\int\bar{\theta}\Gamma_{IJ}d\theta\wedge\Pi^{I}\wedge\Pi^{J}.
\end{equation}
Here, the indices $i,j,k=1,2,3$ label the worldvolume coordinates,
while $I,J,K=0,\dots,10$ denote the spacetime coordinates.

The object $\Pi_{i}^{I}$ is the supervielbein, defined by 
\begin{equation}
\Pi_{i}^{I}=\partial_{i}X^{I}-i\bar{\theta}\Gamma^{I}\partial_{i}\theta,
\end{equation}

where $X^{I}$ are the spacetime coordinates, and $\theta^{\alpha}$
are the fermionic coordinates in superspace. The index $\alpha$ labels
the components of a 32-component Majorana spinor. The conjugate spinor
$\bar{\theta}$ is defined as $\bar{\theta}=\theta^{T}C$, where $C$
is the charge conjugation matrix.

We adopt the following notational conventions:
\begin{equation}
\Pi^{I}\equiv\Pi_{i}^{I}d\sigma^{i},\ d\theta^{\alpha}\equiv\partial_{i}\theta^{\alpha}d\sigma^{i}
\end{equation}

where $\sigma^{i}$ are the worldvolume coordinates.

The gamma matrices $\Gamma_{I}$ satisfy the Clifford algebra
\begin{equation}
[\Gamma_{I},\Gamma_{J}]_{+}=2\eta_{IJ},
\end{equation}

where $\eta_{IJ}$ is the Minkowski metric, and $[A,B]_{+}\equiv AB+BA$.
From the gamma matrices, we define
\begin{equation}
\Gamma_{IJ}\equiv\frac{1}{2}[\Gamma_{I},\Gamma_{J}],
\end{equation}

and, for convenience, we also use the shorthand
\begin{equation}
\Gamma_{i}\equiv\Pi_{i}^{I}\Gamma_{I}.
\end{equation}

Although the full action is invariant under general volume-preserving
coordinate transformations, we adopt a gauge-fixing condition $e=1$,
which reduces this invariance to the subclass of\textbf{ volume-preserving
deformations} (VPD).

\section{Reformulation of the Super M2-Brane Action Using Nambu Brackets}

In this section, we reformulate the supermembrane action using Nambu
brackets.

\LyXZeroWidthSpace{}

\subsection{Reformulation of the Action}

A description of the supermembrane action based on the Nambu bracket
has previously been given in \cite{Kamani_2011}. In the present work,
however, we reconstruct the formulation in a unified exponential form
and extend the theoretical framework to incorporate consistency with
both the RVPD structure and the restricted $\kappa$-symmetry.

\LyXZeroWidthSpace{}

When the gauge fixing $e=1$ is imposed, the action becomes

\begin{equation}
S_{NB}=-\frac{T}{2}\int d^{3}\sigma\left(\epsilon^{ijk}\Pi_{i}^{I}\Pi_{j}^{J}\Pi_{k}^{K}\right)^{2}+1,
\end{equation}

\begin{equation}
S_{WZ}=\frac{i}{3!}\int\bar{\theta}\Gamma_{IJ}d\theta\wedge\Pi^{I}\wedge\Pi^{J}.
\end{equation}

We note the following properties of the spinors:
\begin{equation}
\bar{\theta}\Gamma^{I}\theta=0,\ \bar{\theta}\Gamma^{IJK}\theta=0,\ d\bar{\theta}\Gamma^{IJ}d\theta=0,\ d\bar{\theta}\Gamma^{IJKL}d\theta=0.
\end{equation}

From these, it follows that
\begin{equation}
\partial_{i}\Pi_{i}^{I}=\partial_{i}^{2}X^{I}-i\bar{\theta}\Gamma^{I}\partial_{i}^{2}\theta.
\end{equation}

Using these identities, we can expand the pullback term as

\begin{equation}
\begin{array}{ccc}
\epsilon^{ijk}\Pi_{i}^{I}\Pi_{j}^{J}\Pi_{k}^{K} & = & \{X^{I},X^{J},X^{K}\}\\
 &  & -i3\bar{\theta}^{\alpha}\left(\{\left(\Gamma^{I}\theta\right)_{\alpha},X^{J},X^{K}\}+\{X^{I},\left(\Gamma^{J}\theta\right)_{\alpha},X^{K}\}+\{X^{I},X^{J},\left(\Gamma^{K}\theta\right)_{\alpha}\}\right)\\
 &  & -3\bar{\theta}^{\alpha}\bar{\theta}^{\beta}\left(\{\left(\Gamma^{I}\theta\right)_{\alpha},\left(\Gamma^{J}\theta\right)_{\beta},X^{K}\}+\{\left(\Gamma^{I}\theta\right)_{\alpha},X^{J},\left(\Gamma^{K}\theta\right)_{\beta}\}+\{X^{I},\left(\Gamma^{J}\theta\right)_{\alpha},\left(\Gamma^{K}\theta\right)_{\beta}\}\right)\\
 &  & +\bar{\theta}^{\alpha}\bar{\theta}^{\beta}\bar{\theta}^{\gamma}\{\left(\Gamma^{I}\theta\right)_{\alpha},\left(\Gamma^{J}\theta\right)_{\beta},\left(\Gamma^{K}\theta\right)_{\gamma}\}.
\end{array}
\end{equation}

Similarly, the Wess--Zumino term reduces to
\begin{equation}
\bar{\theta}\Gamma_{IJ}d\theta\wedge\Pi^{I}\wedge\Pi^{J}=\bar{\theta}\{\Gamma_{IJ}\theta,X^{I},X^{J}\}-2i\bar{\theta}^{\alpha}\bar{\theta}^{\beta}\{\left(\Gamma_{IJ}\theta\right)_{\alpha},\left(\Gamma^{I}\theta\right)_{\beta},X^{J}\},
\end{equation}

with the identity
\begin{equation}
\bar{\theta}^{\alpha}\bar{\theta}^{\beta}\bar{\theta}^{\gamma}\{\left(\Gamma_{IJ}\theta\right)_{\alpha},\left(\Gamma^{I}\theta\right)_{\beta},\left(\Gamma^{J}\theta\right)_{\gamma}\}=0
\end{equation}

also holding.

From these expressions, we conclude that the full supermembrane action
can be entirely rewritten in terms of Nambu brackets.

\LyXZeroWidthSpace{}

To simplify notation, we introduce the supersymmetry transformation
symbols
\begin{equation}
\delta_{S,\alpha}X^{I}\equiv i\left(\Gamma^{I}\theta\right)_{\alpha},\ \delta_{S,\alpha}\theta^{\beta}\equiv\delta_{\alpha}^{\beta}.
\end{equation}

Using this notation, the pullback term and the Wess--Zumino term
can be written as
\begin{equation}
\epsilon^{ijk}\Pi_{i}^{I}\Pi_{j}^{J}\Pi_{k}^{K}=e^{\bar{\theta}\delta_{S}}\{X^{I},X^{J},X^{K}\},
\end{equation}

\begin{equation}
\bar{\theta}\Gamma_{IJ}d\theta\wedge\Pi^{I}\wedge\Pi^{J}=\bar{\theta}e^{\bar{\theta}\delta_{S}}\{\Gamma_{IJ}\theta,X^{I},X^{J}\}.
\end{equation}

From this, the total supermembrane action can be fully rewritten in
terms of Nambu brackets as
\begin{equation}
S=S_{NB}+S_{WZ},
\end{equation}

\begin{equation}
S_{NB}=-\frac{T}{2}\int d^{3}\sigma\left(e^{\bar{\theta}\delta_{S}}\{X^{I},X^{J},X^{K}\}\right)^{2},
\end{equation}

\begin{equation}
S_{WZ}=\frac{i}{3!}\int d^{3}\sigma\bar{\theta}e^{\bar{\theta}\delta_{S}}\{\Gamma_{IJ}\theta,X^{I},X^{J}\}.
\end{equation}

Thus, the entire supermembrane action is completely expressed in terms
of Nambu brackets.\footnote{In the expression $\bar{\theta}e^{\bar{\theta}\delta_{S}}\{\Gamma_{IJ}\theta,X^{I},X^{J}\}$,
the $\bar{\theta}$ outside the exponential is not part of the Nambu
bracket. It serves as a prefactor and should not be confused with
the operator $\bar{\theta}\delta_{S}$ inside the exponent.}

\subsection{Decomposition into Poisson Brackets}

As in the bosonic case, the Nambu bracket can be decomposed into Poisson
brackets as follows:
\begin{equation}
\{A,B,C\}=\{\tau(A,B),C\}+\frac{\partial C}{\partial\sigma^{3}}\{A,B\}+\Sigma(A,B;C)
\end{equation}

where
\begin{itemize}
\item $\tau(A,B)\equiv\partial_{\sigma^{3}}AB-\partial_{\sigma^{3}}BA,$
\item $\{A,B\}\equiv\epsilon^{ab}\frac{\partial A}{\partial\sigma^{a}}\frac{\partial B}{\partial\sigma^{b}},\ a,b=1,2$,
\item $\Sigma(A,B;C)\equiv A\{\partial_{\sigma^{3}}B,C\}-B\{\partial_{\sigma^{3}}A,C\}$.
\end{itemize}
This decomposition is an identity that rewrites the Nambu bracket
in an equivalent form. In the matrix regularization scheme, the Poisson
bracket $\{A,B\}$ is then replaced by (anti)commutators

\begin{equation}
[A,B]=AB-BA,
\end{equation}

and the Nambu bracket $\{A,B,C\}$ is replaced by the triple commutator

\begin{equation}
[A,B,C]=[\tau(A,B),C]+\frac{\partial C}{\partial\sigma^{3}}[A,B]+\Sigma(A,B;C),
\end{equation}

which is totally antisymmetric in $(A,B,C)$.

Note that the invariance of the action under volume-preserving deformations
(VPD), given by $\delta X^{I}=\{Q_{1},Q_{2},X^{I}\}$ and $\delta\theta=\{Q_{1},Q_{2},\theta\}$,
was already discussed in Section 2. We shall continue to make use
of this symmetry structure in our formulation.

\subsection{$\kappa$-Symmetry}

In addition to supersymmetry and volume-preserving deformations (VPD),
the action also possesses a local fermionic symmetry known as $\kappa$-symmetry.
This symmetry was first discussed by \cite{de_Azc_rraga_1982,Siegel_1983},
and was incorporated into string theory by \cite{Green_1984}.

The $\kappa$-symmetry transformations are given by:

\begin{equation}
\delta_{\kappa}\theta=(1+\Gamma)\kappa(\sigma),
\end{equation}

\begin{equation}
\delta_{\kappa}X^{I}=i\bar{\theta}\Gamma^{I}\delta_{\kappa}\theta,
\end{equation}

\begin{equation}
\delta_{\kappa}\Pi_{i}^{I}=2i\bar{\kappa}\Gamma^{I}\partial_{i}\theta.
\end{equation}

Here, $\Gamma$ is called the chiral operator, which satisfies $\Gamma^{2}=1$.
Therefore, the projection operator $(1+\Gamma)$ acts on only half
of the components of $\kappa$, effectively reducing the fermionic
degrees of freedom by half.

\LyXZeroWidthSpace{}

The chiral operator $\Gamma$ is defined as

\begin{equation}
\Gamma\equiv\frac{1}{6\sqrt{-g}}\epsilon^{ijk}\Pi_{i}^{I}\Pi_{j}^{J}\Pi_{k}^{K}\Gamma_{IJK},
\end{equation}

where
\begin{equation}
\Gamma_{IJK}\equiv\Gamma_{[I}\Gamma_{J}\Gamma_{K]}
\end{equation}

denotes the totally antisymmetrized product of gamma matrices.

\section{RVPD Gauge Restriction and $\kappa$-Symmetry}

In this section, we examine how the gauge restriction introduced in
our previous work on the bosonic membrane model \cite{https://doi.org/10.48550/arxiv.2504.05940}
affects the structure of $\kappa$-symmetry in the supersymmetric
extension.

\LyXZeroWidthSpace{}

The gauge restriction we introduced in the bosonic membrane model\cite{https://doi.org/10.48550/arxiv.2504.05940}
is given by:
\begin{equation}
C_{I}\partial_{\sigma^{3}}X^{I}=\sigma^{3}.
\end{equation}

When we apply a supersymmetry transformation to this condition, we
obtain
\begin{equation}
C_{I}\Gamma^{I}\partial_{\sigma^{3}}\theta^{\alpha}=0.
\end{equation}

Integrating this with respect to $\sigma^{3}$, we find
\begin{equation}
C_{I}\left(\Gamma^{I}\theta\right)_{\alpha}=f_{\alpha}(\sigma^{1},\sigma^{2}),
\end{equation}

where $f_{\alpha}$ is an arbitrary function of $\sigma^{1}$ and
$\sigma^{2}$.

Next, we consider applying a $\kappa$-transformation to the original
gauge restriction:
\begin{equation}
C_{I}\partial_{\sigma^{3}}\delta_{\kappa}X^{I}=C_{I}\partial_{\sigma^{3}}\left(\bar{\theta}\Gamma^{I}\delta_{\kappa}\theta\right).
\end{equation}

Substituting the previous result, this becomes

\begin{equation}
C_{I}\bar{f}(\sigma_{1},\sigma_{2})\partial_{\sigma^{3}}\delta_{\kappa}\theta=0.
\end{equation}

If this condition is to hold for arbitrary $f_{\alpha}$, then we
must require 
\begin{equation}
\partial_{\sigma^{3}}(1+\Gamma)\kappa=0.
\end{equation}

Integrating over $\sigma^{3}$, we obtain
\begin{equation}
(1+\Gamma)\kappa(\sigma_{1},\sigma_{2},\sigma_{3})=\tilde{\kappa}(\sigma_{1},\sigma_{2}).
\end{equation}

This implies that, under the gauge restriction, the $\kappa$-symmetry
is reduced to a restricted form:
\begin{equation}
\delta_{\tilde{\kappa}}\theta=\tilde{\kappa}(\sigma_{1},\sigma_{2}),
\end{equation}
\begin{equation}
\delta_{\tilde{\kappa}}X^{I}=i\bar{\theta}\Gamma^{I}\tilde{\kappa}.
\end{equation}

Applying this restricted $\tilde{\kappa}$-transformation twice yields:
\begin{equation}
\delta_{\tilde{\kappa}_{2}}\delta_{\tilde{\kappa}_{1}}\theta=0,
\end{equation}
\begin{equation}
\delta_{\tilde{\kappa}_{2}}\delta_{\tilde{\kappa}_{1}}X^{I}=i\bar{\tilde{\kappa}}_{2}\Gamma^{I}\tilde{\kappa}_{1}.
\end{equation}

Continuing from the above, recall that the RVPD transformation acts
on the fields as:\footnote{Here $a,b\in\{1,2\}$ and $\epsilon^{12}=1$.}
\begin{equation}
\delta_{R(Q_{1},Q_{2})}X^{I}=\epsilon^{ab}\partial_{a}\tau(Q_{1},Q_{2})\partial_{b}X^{I},
\end{equation}

where the charges $Q_{1}$ and $Q_{2}$ satisfy the following constraints:
\begin{equation}
\epsilon^{ab}\partial_{a}Q_{1}\partial_{b}Q_{2}=0,
\end{equation}

\begin{equation}
\partial_{a}\partial_{\sigma^{3}}Q_{1,2}=0.
\end{equation}

We now demonstrate that the RVPD transformation can be generated as
a composite effect of two restricted $\tilde{\kappa}$-transformations
with appropriately chosen parameters.

To this end, we introduce a fermionic (Grassmann-odd) variable $\xi$
, and define the restricted $\kappa$-parameters as
\begin{equation}
\tilde{\kappa}_{1}=\left(\frac{\partial}{\partial\xi}+\xi\left(\partial_{\sigma^{3}}Q_{1}\epsilon^{ab}\partial_{a}Q_{2}\partial_{b}g\right)\right)\psi_{1},
\end{equation}

\begin{equation}
\tilde{\kappa}_{2}=\left(\frac{\partial}{\partial\xi}-\xi\left(\partial_{\sigma^{3}}Q_{2}\epsilon^{ab}\partial_{a}Q_{1}\partial_{b}g\right)\right)\psi_{2},
\end{equation}

where $\psi_{1}$ and $\psi_{2}$ are constant spinors to be normalized
later, and $g(\sigma_{1},\sigma_{2},\sigma_{3})$ is an arbitrary
function of the worldvolume coordinates. The successive action of
these restricted $\kappa$-transformations on $X^{I}$ yields

\begin{equation}
\delta_{\tilde{\kappa}_{2}}\delta_{\tilde{\kappa}_{1}}X^{I}=i\bar{\tilde{\kappa}}_{2}\Gamma^{I}\tilde{\kappa}_{1}=i\epsilon^{ab}\partial_{a}\tau(Q_{1},Q_{2})\partial_{b}g(\bar{\psi}_{2}\Gamma^{I}\psi_{1}).
\end{equation}

This can be rewritten as

\begin{equation}
\delta_{\tilde{\kappa}_{2}}\delta_{\tilde{\kappa}_{1}}X^{I}=i(\bar{\psi}_{2}\Gamma^{I}\psi_{1})C_{J}\epsilon^{ab}\partial_{a}\tau(Q_{1},Q_{2})\partial_{b}X^{J},
\end{equation}

where $g$ is taken to be linear in $X^{J}$ as $g=C_{I}X^{I}$.

Here, we introduce the projection operator
\begin{equation}
P_{\ J}^{I}\equiv(\bar{\psi}_{2}\Gamma^{I}\psi_{1})C_{J}.
\end{equation}

We can compute its square:
\begin{equation}
P_{\ J}^{I}P_{\ K}^{J}=(\bar{\psi}_{2}\Gamma^{I}\psi_{1})C_{J}(\bar{\psi}_{2}\Gamma^{J}\psi_{1})C_{K}.
\end{equation}

If we normalize the spinors so that
\begin{equation}
C_{J}(\bar{\psi}_{2}\Gamma^{J}\psi_{1})=1,
\end{equation}

then the operator becomes idempotent:
\begin{equation}
P_{\ J}^{I}P_{\ K}^{J}=P_{\ K}^{I}.
\end{equation}

Therefore, we can write
\begin{equation}
\delta_{\tilde{\kappa}_{2}}\delta_{\tilde{\kappa}_{1}}X^{I}=i\epsilon^{ab}\partial_{a}\tau(Q_{1},Q_{2})\partial_{b}\left(P_{\ J}^{I}X^{J}\right).
\end{equation}

This result shows that by choosing appropriate parameters for the
restricted $\kappa$-symmetry, the composite transformation acts as
an RVPD transformation. In this sense, the RVPD algebra can be embedded
within the algebra of restricted $\kappa$-transformations.

\LyXZeroWidthSpace{}

To summarize, we have reformulated the supermembrane action in terms
of Nambu brackets, and further decomposed it into Poisson brackets
involving only the directions $i=1,2.$

\LyXZeroWidthSpace{}

By examining the supersymmetric extension of the gauge restriction
condition, we found that the original $\kappa$-symmetry is reduced
to a restricted form $\tilde{\kappa}$, and that this restricted symmetry
forms a closed algebra with the RVPD transformations.

\LyXZeroWidthSpace{}

Therefore, within a matrix model constructed by replacing the Poisson
brackets with (anti)commutators, this algebra can be used to define
and analyze BPS conditions. As a result, one can systematically study
the supersymmetric matrix model for membranes, and evaluate the stability
of the solutions previously found in the bosonic case.

\section{BPS Analysis of Solutions}

In the bosonic matrix model for M2-branes, a variety of classical
solutions have been found\cite{https://doi.org/10.48550/arxiv.2504.05940}.
When we set $\theta$ = 0, these also serve as classical configurations
in the supersymmetric membrane theory. In this section, we examine
whether these configurations satisfy appropriate BPS conditions.

\LyXZeroWidthSpace{}

A systematic analysis of BPS projections for M2-branes was provided
in \cite{Townsend-1}. Here, however, we perform the analysis using
the restricted $\kappa$-symmetry $\tilde{\kappa}$ constructed under
the RVPD gauge restriction. This approach allows us to directly assess
the amount of preserved supersymmetry associated with each solution.

\subsection{Chiral Operator and Its Non-Idempotency}

At this point, we must pay particular attention to the chiral operator
$\Gamma$, which plays a crucial role in defining the BPS projection
condition. In this paper, we choose the normalization factor a on
a case-by-case basis so as to satisfy $\Gamma^{2}=1$ whenever possible.

In our formalism, it is defined as:
\begin{equation}
\Gamma\equiv\frac{1}{a}e^{\bar{\theta}\delta_{S}}[X^{I},X^{J},X^{K}]\Gamma_{IJK},
\end{equation}

where the normalization factor $a$ will be fixed later, and $\Gamma_{IJK}=\Gamma_{[I}\Gamma_{J}\Gamma_{K]}$
denotes the totally antisymmetrized product of gamma matrices.

Since we are considering classical solutions with $\theta=0$, the
exponential factor drops out and we have:
\begin{equation}
\Gamma=\frac{1}{a}[X^{I},X^{J},X^{K}]\Gamma_{IJK}.
\end{equation}

To analyze the BPS projection condition, we must examine whether $\Gamma^{2}=1$
holds. The square of the chiral operator becomes:
\begin{equation}
\Gamma^{2}=[X^{I},X^{J},X^{K}][X^{L},X^{M},X^{N}]\Gamma_{IJK}\Gamma_{LMN}.
\end{equation}

The product of gamma matrices can be expanded using standard Clifford
algebra identities\footnote{The product of antisymmetrized gamma matrices can be expanded using
the standard Clifford-algebra identity (see e.g. A. Van Proeyen, Tools
for supersymmetry\cite{https://doi.org/10.48550/arxiv.hep-th/9910030},
eq. (3.41)). Specializing it to $i=j=3$ and adopting the antisymmetrization
convention of Park (\cite{Park_2022}, eq. (2.4)).}:
\begin{equation}
\Gamma_{IJK}\Gamma_{LMN}=\Gamma_{IJKLMN}-9\eta_{I[L}\Gamma_{JK]MN}+18\eta_{I[L}\eta_{JM}\Gamma_{K]N}-6\eta_{I[L}\eta_{JM}\eta_{KN]},
\end{equation}

which leads to the expansion:
\begin{equation}
\begin{aligned}\Gamma^{2}= & -6[X^{I},X^{J},X^{K}][X_{I},X_{J},X_{K}]\\
 & +18[X^{I},X^{J},X^{P}][X^{K},X^{L},X_{P}]\Gamma_{IJKL}\\
 & +\frac{1}{6}[X^{[I},X^{J},X^{K}][X^{L},X^{M},X^{N]}]\Gamma_{IJKLMN}.
\end{aligned}
\end{equation}

To proceed further, we decompose the triple commutator as
\begin{equation}
[X^{I},X^{J},X^{K}]=[\tau(X^{I},X^{J}),X^{K}]+\frac{\partial X^{K}}{\partial\sigma^{3}}[X^{I},X^{J}]+\Sigma(X^{I},X^{J};X^{K})
\end{equation}

with 
\begin{itemize}
\item $\tau(A,B)\equiv\partial_{\sigma^{3}}AB-\partial_{\sigma^{3}}BA$,
\item $[A,B]=AB-BA$,
\item $\Sigma(A,B;C)=A[\partial_{\sigma^{3}}B,C]-B[\partial_{\sigma^{3}}A,C]$.
\end{itemize}
Substituting this decomposition into the expression for $\Gamma^{2}$
, one finds terms of the form:
\begin{equation}
[X^{I},X^{J},X^{K}][X^{L},X^{M},X^{N}]\Gamma_{IJKLMN}=\partial_{\sigma^{3}}X^{I}X^{J}X^{K}\partial_{\sigma^{3}}X^{L}X^{M}X^{N}\Gamma_{IJKLMN}+\dots.
\end{equation}

Now, since we are considering noncommutative membrane solutions, we
assume that the two-bracket $[X^{I},X^{J}]$ is given by a constant
antisymmetric tensor:
\begin{equation}
[X^{I},X^{J}]=c^{IJ}\ \mathrm{or}\ 0.
\end{equation}

Thus, we find that the leading contribution to $\Gamma^{2}$ is of
order $\mathcal{O}([X^{I},X^{J}])$, and the idempotency condition
$\Gamma^{2}=1$ is generally not satisfied unless we take a continuum
limit or consider specific configurations. Therefore, the BPS projection
condition must be analyzed on a case-by-case basis depending on the
background solution.

\subsection{Particle-like Solution}

We begin by considering a particle-like solution given by:

\begin{equation}
X^{0}=\sigma^{3},
\end{equation}

\begin{equation}
X^{1,\dots,10}=f^{1,\dots,10}(\sigma^{3}),
\end{equation}

\begin{equation}
\theta^{\alpha}=0.
\end{equation}

For this configuration, the fermionic variation $\delta_{\epsilon}\theta^{\alpha}$
is examined. According to the discussion in the previous section,
the supersymmetry parameter can be canceled by the restricted $\kappa$-symmetry:
\begin{equation}
\epsilon=-\tilde{\kappa}(\sigma_{1},\sigma_{2})=(1+\Gamma)\kappa(\sigma_{1},\sigma_{2},\sigma_{3}).
\end{equation}

This shows that the supersymmetry variation is gauge equivalent to
zero under the restricted $\kappa$-transformation.

\LyXZeroWidthSpace{}

Now, recall that the chiral operator $\Gamma$ is given by:
\begin{equation}
\Gamma=\frac{1}{a}e^{\bar{\theta}\delta_{S}}[X^{I},X^{J},X^{K}]\Gamma_{IJK}.
\end{equation}

We set $a=1$ for simplicity. Since we are evaluating the solution
at $\theta=0$, the exponential factor becomes trivial. Furthermore,
the non-vanishing components of the derivatives are:
\begin{equation}
\partial_{\sigma^{3}}X^{0}=1,
\end{equation}

\begin{equation}
\partial_{\sigma^{3}}X^{1,\dots,10}=\partial_{\sigma^{3}}f^{1,\dots10}.
\end{equation}

As this configuration involves only commuting scalar functions of
$\sigma^{3}$ , all commutators vanish:

\begin{equation}
[X^{I},X^{J},X^{K}]=0\Rightarrow\Gamma=0.
\end{equation}

Now consider the variation of $X^{I}$ under supersymmetry and $\kappa$-symmetry:
\begin{equation}
\delta_{\epsilon}X^{I}+\delta_{\kappa}X^{I}=i\bar{\theta}\Gamma^{I}(\epsilon+\tilde{\kappa})=0,
\end{equation}

which is trivially satisfied because $\theta=0$.

Therefore, this background preserves all supersymmetries that can
be canceled by $\tilde{\kappa}$, amounting to half of the original
32 components. Hence, the configuration corresponds to a $1/2$-BPS
state, preserving $16$ supersymmetries.

\subsection{Noncommutative Membrane Solution}

Next, we consider a noncommutative membrane solution defined by:

\begin{equation}
X^{0}=\sigma^{3},
\end{equation}

\begin{equation}
[X^{1},X^{2}]=i,
\end{equation}

\begin{equation}
X^{3,\dots,10}=0,
\end{equation}

\begin{equation}
\theta=0.
\end{equation}

Since $\theta=0$, the fermionic variation is again canceled by the
restricted $\kappa$-symmetry:
\begin{equation}
\epsilon=-\tilde{\kappa}.
\end{equation}

Accordingly, the variation of $X^{I}$ vanishes automatically as in
the particle-like solution.

Now, let us compute the chiral operator $\Gamma$ , which is defined
by:
\begin{equation}
\Gamma=e^{\bar{\theta}\delta_{S}}[X^{I},X^{J},X^{K}]\Gamma_{IJK}.
\end{equation}

For $\theta=0$ , the exponential becomes trivial, and we have:
\begin{equation}
\Gamma=[X^{I},X^{J},X^{K}]\Gamma_{IJK}.
\end{equation}

In the present solution, the nonvanishing components are:
\begin{equation}
\partial_{\sigma^{3}}X^{0}=1,
\end{equation}

\begin{equation}
[X^{1},X^{2}]=i,
\end{equation}

while all others vanish. Therefore, the triple commutator reduces
to

\begin{equation}
[X^{I},X^{J},X^{K}]=\frac{\partial X^{K}}{\partial\sigma^{3}}[X^{I},X^{J}],
\end{equation}

and the chiral operator becomes:
\begin{equation}
\Gamma=\left(\frac{\partial X^{K}}{\partial\sigma^{3}}[X^{I},X^{J}]\right)\Gamma_{IJK}=\Gamma_{012}.
\end{equation}

Thus, the supersymmetry parameter satisfies:
\begin{equation}
\epsilon=-\tilde{\kappa}=(1+\Gamma_{012})\kappa.
\end{equation}

Multiplying both sides by $(1-\Gamma_{012}$) from the left, we obtain
the projection condition: 
\begin{equation}
\epsilon=\Gamma_{012}\epsilon.
\end{equation}

This projects out half of the remaining supersymmetries. Since the
original theory has $32$ supercharges, this solution preserves $8$
components. Hence, it corresponds to a $1/4$-BPS state.

\subsection{Multiple Membrane Configuration}

In the case of multiple membrane solutions, the structure is essentially
the same: each $(X^{1},X^{2})$ pair satisfies a commutation relation
of the form 

\begin{equation}
[X^{1},X^{2}]=c^{12},
\end{equation}

where $c^{12}$ is a constant matrix (typically diagonal). Since this
structure is identical to the single-membrane case with respect to
the $(X^{1},X^{2})$ sector, the chiral operator remains

\begin{equation}
\Gamma=\Gamma_{012}.
\end{equation}

Therefore, the same projection condition $\epsilon=\Gamma_{012}\epsilon$
applies, and the preserved supersymmetries are again reduced to $8$
components out of $32$.

As a result, the multiple membrane solution also corresponds to a
$1/4$-BPS state, identical to the single noncommutative membrane
configuration discussed in the previous subsection.

\subsection{Membrane Solutions in $4,6,8,$ and $10$ Dimensions}

We now consider extended membrane solutions in higher dimensions.

\subsubsection{4-dimensional membrane:}

The 4-dimensional configuration is defined by

\begin{equation}
X^{0}=\sigma^{3},
\end{equation}

\begin{equation}
[X^{1},X^{2}]=i,
\end{equation}

\begin{equation}
[X^{3},X^{4}]=i.
\end{equation}

The supersymmetry parameter is again given by
\begin{equation}
\epsilon=-\tilde{\kappa}=(1+\Gamma)\kappa.
\end{equation}

The chiral operator becomes
\begin{equation}
\Gamma=\frac{1}{a}[X^{I},X^{J},X^{K}]\Gamma_{IJK}=\frac{1}{a}\left(\frac{\partial X^{K}}{\partial\sigma^{3}}[X^{I},X^{J}]\right)\Gamma_{IJK}=\frac{1}{a}\left(\Gamma_{012}+\Gamma_{034}\right).
\end{equation}

To ensure that $\Gamma^{2}=1$, we choose the normalization $1/a=1/\sqrt{2}$
. With this, the supersymmetry parameter becomes:
\begin{equation}
\epsilon=(1+\frac{i}{\sqrt{2}}\left(\Gamma_{012}+\Gamma_{034}\right))\kappa.
\end{equation}

In order for this condition to be satisfied, the following independent
projection conditions must hold:

\begin{equation}
(1-\Gamma_{012})\epsilon=0,
\end{equation}

\begin{equation}
(1-\Gamma_{034})\epsilon=0.
\end{equation}

Each projection reduces the number of preserved supercharges by a
factor of $1/2$. Thus, the number of preserved components becomes:

\begin{equation}
32\times\left(\frac{1}{2}\right)^{3}=4.
\end{equation}

This configuration therefore corresponds to a $1/8$-BPS state.

\subsubsection{6- and 8-dimensional membranes:}

In the same way, we can construct higher-dimensional membrane solutions
by adding additional commuting pairs. For instance:
\begin{itemize}
\item In the 6-dimensional case, adding $[X^{5},X^{6}]=i$ imposes a third
projection condition, reducing the preserved supersymmetries to
\begin{equation}
32\times\left(\frac{1}{2}\right)^{4}=2,
\end{equation}
corresponding to a $1/16$-BPS state.
\item In the 8-dimensional case, with four independent noncommuting pairs(e.g.,
$[X^{7},X^{8}]=i$), we obtain:
\begin{equation}
32\times\left(\frac{1}{2}\right)^{5}=1,
\end{equation}
leading to a 1/32-BPS state.
\end{itemize}

\subsubsection{10-dimensional membrane:}

In the 10-dimensional case, if all five pairs $[X^{2n-1},\ X^{2n}]=i$
are activated, the projection would formally require:

\begin{equation}
\left(1-\Gamma_{0AB}\right)\epsilon=0
\end{equation}

for five independent $\Gamma_{0AB}$, which would overconstrain $\epsilon$,
leaving no supersymmetry preserved. Therefore, the 10-dimensional
membrane configuration does not correspond to a BPS state and is expected
to be unstable.

\section{Summary and Discussion}

By considering the supersymmetric extension of the gauge restriction
condition, we have shown that the $\kappa$-symmetry is reduced to
a restricted form, $\tilde{\kappa}$, under this constraint. Remarkably,
this restricted $\kappa$-symmetry forms a closed algebra together
with the Restricted Volume-Preserving Deformations (RVPD).

\LyXZeroWidthSpace{}

This algebraic structure enables us to systematically formulate and
analyze BPS conditions within a matrix model framework for supermembranes.
In particular, it allows us to assess the stability of various classical
solutions obtained in the bosonic matrix model proposed in \cite{https://doi.org/10.48550/arxiv.2504.05940}.

\LyXZeroWidthSpace{}

Here we list the preserved supercharges and the corresponding BPS
fractions for each configuration.

The classification of BPS states is summarized in Table \ref{tab:Classification-of-BPS}.

\begin{table}[h]
\centering
\begin{tabular}{|c|c|c|}
\hline 
Configuration & Number of Preserved Supercharges & BPS Fraction\tabularnewline
\hline 
\hline 
Particle-like solution & 16 & 1/2-BPS\tabularnewline
\hline 
(Multiple) Noncommutative membrane & 8 & 1/4-BPS\tabularnewline
\hline 
4-dimensional membrane & 4 & 1/8-BPS\tabularnewline
\hline 
6-dimensional membrane & 2 & 1/16-BPS\tabularnewline
\hline 
8-dimensional membrane & 1 & 1/32-BPS\tabularnewline
\hline 
10-dimensional membrane & 0 & unstable\tabularnewline
\hline 
\end{tabular}

\caption{Classification of BPS states obtained in the supersymmetric extension
of the Lorentz-covariant matrix model under the RVPD constraint in
the supersymmetric matrix model, based on the $\tilde{\kappa}$-symmetry
analysis. The preserved supercharges are counted out of the total
32 supersymmetries in eleven-dimensional supergravity.\label{tab:Classification-of-BPS}}

\end{table}

These results demonstrate that the supersymmetric matrix model for
membranes accommodates a rich spectrum of BPS configurations with
varying degrees of supersymmetry.

\LyXZeroWidthSpace{}

This work has been conducted entirely at the classical level. The
next crucial step is to investigate the quantization of the model
and understand how these classical BPS structures manifest in the
quantum theory. In particular, the RVPD--$\tilde{\kappa}$ algebra
may provide a natural starting point for defining the physical Hilbert
space, while matrix regularization offers a concrete framework for
path-integral or canonical quantization. We expect that the discrete
BPS spectrum, its stability under quantum corrections, and the possible
lifting or splitting of degenerate states can be systematically analyzed
within this setting. Questions regarding the stability, spectrum,
and moduli of quantum BPS states remain open and are important directions
for future research.

\LyXZeroWidthSpace{}

Furthermore, the noncommutative membrane solutions constructed in
our matrix model exhibit structural parallels to D2-branes in type
IIA string theory. Upon dimensional reduction along a compactified
$11$th direction, the configuration $[X^{1},X^{2}]=i$ directly maps
to the noncommutative gauge field background on a D2-brane. In this
sense, the present Lorentz covariant matrix model can be regarded
as a natural lift of the D2-brane matrix description to M-theory,
with the RVPD symmetry playing a role analogous to the gauge invariance
in D-brane worldvolume theories.

\LyXZeroWidthSpace{}

It is worth noting that our formalism, based on the Nambu bracket
and RVPD gauge constraints, may lend itself naturally to generalizations
beyond M2-branes. In particular, it would be intriguing to explore
whether a 4-bracket or 5-bracket version of the current framework
could provide a matrix model description of M5-branes, potentially
capturing the dynamics of self-dual 3-form fields in a nonperturbative
setting. Although such an extension involves significant challenges---both
algebraic and geometric---it remains an important direction for future
investigation.

\section*{Acknowledgments}

I would like to express my sincere gratitude to Akio Sugamoto for
his careful reading of the manuscript and for providing valuable comments.
I am also indebted to Shiro Komata for his thorough review of the
final draft and for offering detailed suggestions.

\bibliographystyle{unsrt}
\bibliography{SuperRVPD_paper_katagiri}

\end{document}